\documentclass{PoS}

\title{Radiative corrections to semileptonic decay rates}

\ShortTitle{Radiative corrections to semileptonic decay rates}

\author{\speaker{C.\,T.\,Sachrajda}\,\thanks{cts@soton.ac.uk}\\
        Department of Physics and Astronomy, University of Southampton, Southampton SO17 1BJ, UK}

\author{M.\,Di Carlo, G.\,Martinelli,\\
        Dip. di Fisica and INFN Sezione di Roma La Sapienza, Piazzale Aldo Moro 5, 00185 Roma, Italy}

\author{D.\,Giusti, V.\,Lubicz,\\
        Dip. di Matematica e Fisica, University of   Roma Tre and INFN   Roma Tre,\\ Via della Vasca Navale 84, I-00146 Rome, Italy}

\author{F.\,Sanfilippo, S.\,Simula,\\
        INFN   Roma Tre, Via della Vasca Navale 84, I-00146 Rome, Italy}
        
\author{N.\,Tantalo,\\
        University of Rome Tor Vergata and INFN Roma Tor Vergata,
Via della Ricerca Scientifica 1, I-00133, Rome, Italy }

\abstract{We discuss the theoretical framework required for the computation of radiative corrections to semileptonic decay rates in lattice simulations, and in particular to those for $K_{\ell3}$ decays. This is an extension of the framework we have developed and successfully implemented for leptonic decays. New issues which arise for semileptonic decays, include the presence of unphysical terms which grow exponentially with the time separation between the insertion of the weak Hamiltonian and the sink for the final-state meson-lepton pair. Such terms must be identified and subtracted. We discuss the cancellation of infrared divergences and show that, with the QED$_\mathrm{\,L}$ treatment of the zero mode in the photon propagator, the $O(1/L)$ finite-volume corrections are ``universal". These corrections however, depend not only on the semileptonic form factors $f^\pm(q^2)$ but also on their derivatives $df^\pm/dq^2$. (Here $q$ is the momentum transfer between the initial and final state mesons.) We explain the perturbative calculation which would need to be performed to subtract the $O(1/L)$ finite-volume effects.}

\FullConference{37th International Symposium on Lattice Field Theory - Lattice2019\\
		16-22 June 2019\\
		Wuhan, China}

\begin{document}

\section{Introduction}

The precision of lattice QCD computations of leptonic and semileptonic decay amplitudes has now reached the sub-percent level\,\cite{Aoki:2019cca}. This implies that 
isospin-breaking effects, including electromagnetism, must be included for further progress to be made in the determination of the corresponding CKM matrix elements and other tests of the Standard Model.
When studying radiative corrections to leptonic decays of pseudoscalar mesons at $O(\alpha_\mathrm{em})$, the presence of infrared divergences requires us to consider the rates for both the processes $P\to\ell\bar\nu_{\ell}$ and $P\to\ell\bar\nu_{\ell}\gamma$, which we denote by $\Gamma_0(P\to\ell\bar\nu)$ and $\Gamma_1(P\to\ell\bar\nu)$ respectively, where the subscript {\footnotesize 0,1} denotes the number of photons in the final state. Our initial proposal was to restrict the energy of the final-state photon to be sufficiently small ($E_\gamma<\Delta E_\gamma\simeq 20$\,MeV say) for the dependence on the structure of the meson to be negligible and yet to be within the experimental resolution\,\cite{Carrasco:2015xwa}. It is then convenient to organise the calculation in the form
\begin{equation}\label{eq:masterleptonic}
\Gamma_0+\Gamma_1(\Delta E_\gamma)=\lim_{V\to\infty}\big(\Gamma_0-\Gamma_0^\mathrm{pt}\big)+\lim_{V\to\infty}\big(\Gamma_0^\mathrm{pt}+\Gamma_1(\Delta E_\gamma)\big)\,,\end{equation}
where ``pt" implies that the meson $P$ is treated as being \emph{point-like}.
Each of the two terms on the right-hand side of Eq.\,(\ref{eq:masterleptonic}) is infrared finite and the second term can be calculated in perturbation theory and this was done in Ref.\,\cite{Carrasco:2015xwa}. On the other hand, $\Gamma_0$ must be computed in a lattice simulation, as the amplitude at $O(\alpha_\mathrm{em})$ includes a virtual photon which must be summed over all momenta.

The introduction of the soft energy cut-off $\Delta E_\gamma$ can be avoided by computing amplitudes with a real photon in the final state. Such calculations are now in progress as reported at this conference\,\cite{Kane:2019jtj,deDivitiis:2019uzm}. The non-perturbative evaluation of $\Gamma_1$ has the important practical implication that the method can be applied to the decays of heavy mesons. For example, since $m_{B^\ast}-m_B\simeq 45$\,MeV, the hyperfine splitting for heavy mesons provides another small scale, which limits the scope and precision of the perturbative calculations for soft photons.

\section{Semileptonic decays}

\begin{figure}[t]
\begin{center}
\includegraphics[width=0.25\hsize]{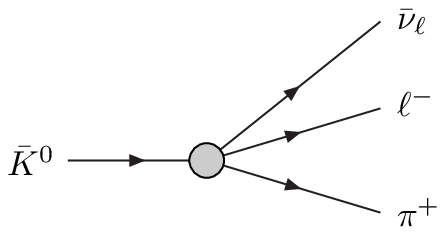}\hspace{0.7in}
\includegraphics[width=0.25\hsize]{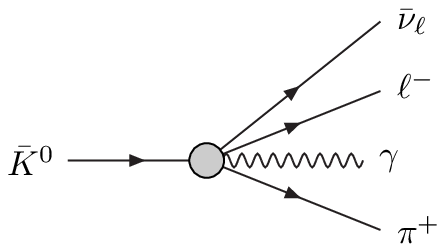}
\end{center}
\caption{Radiative corrections to semileptonic $K_{\ell 3}$ decays at $O(\alpha_\mathrm{em})$ require the evaluation of the rates for both the processes $K\to\pi\ell\bar\nu_{\ell}$ and $K\to\pi\ell\bar\nu_{\ell}\gamma$; the corresponding amplitudes are sketched schematically here.
\label{fig:SL1}}
\end{figure}
For the remainder of this talk, we consider the extension of the ideas of Ref.\,\cite{Carrasco:2015xwa} to semileptonic decays, focussing on $K_{\ell 3}$ decays as illustrated in Fig.\,\ref{fig:SL1}, but noting that the discussion is more general. A particularly appropriate measurable quantity to consider is 
$\frac{d^2\Gamma}{dq^2 ds_{\pi\ell}}$, 
where $q^2=(p_K-p_\pi)^2$ and $s_{\pi\ell}=(p_\pi+p_\ell)^2$.
Following the same procedure as for leptonic decays we write:
\begin{equation}
\frac{d^2\Gamma}{dq^2 ds_{\pi\ell}}=\lim_{V\to\infty}
\left(
\frac{d^2\Gamma_0}{dq^2 ds_{\pi\ell}}
-\frac{d^2\Gamma_0^{\mathrm{pt}}}{dq^2 ds_{\pi\ell}}\right)
+\lim_{V\to\infty}\left(
\frac{d^2\Gamma_0^{\mathrm{pt}}}{dq^2 ds_{\pi\ell}}+\frac{d^2\Gamma_1(\Delta E_\gamma)}{dq^2 ds_{\pi\ell}}\right)\,,\label{eq:SL1}
\end{equation}
where again ``pt" denotes \emph{pointlike} and the infrared divergences 
cancel separately in each of the two terms on the right-hand side. In Eq.\,(\ref{eq:SL1}) we have introduced the soft cut-off $\Delta E_\gamma$ on the energy of the photon, but this can be avoided by computing the amplitudes non-perturbatively with a real final-state photon non-perturbatively. We now discuss a number of issues which arise when considering semileptonic decays which are absent for leptonic decays.

\begin{figure}[t]
\begin{center}
\includegraphics[width=0.35\hsize]{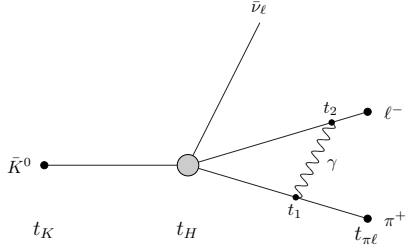}
\end{center}
\caption{Diagram contributing to the $K\to\pi\ell\bar{\nu}_\ell$ correlation function, illustrating the presence of unphysical terms which grow exponentially in time (see text). \label{fig:SL2}}
\end{figure}

\subsection{The presence of unphysical terms which grow exponentially in time.}Consider for illustration the diagram in Fig.\,\ref{fig:SL2}. The integration over the times $t_{1,2}$ yields terms in the momentum sum which are proportional to 
$e^{-(E_{\pi\ell}^\mathrm{int}-E_{\pi\ell}^\mathrm{ext})(t_{\pi\ell}-t_H)}$, where $
E_{\pi\ell}^\mathrm{int}$ and $E_{\pi\ell}^\mathrm{ext}$ are the internal and external energies of the pion-lepton pair and $t_{\pi\ell}$ and $t_H$ are the times of the insertion of the pion-lepton sink and of the weak Hamiltonian $H$. Depending on the choice of the momenta of the final-state pion and lepton, it is possible that the exchange of the photon with an allowed finite-volume  momentum can result in the internal energy being smaller than the external one, $E_{\pi\ell}^\mathrm{int}<E_{\pi\ell}^\mathrm{ext}$, leading to unphysical terms which grow exponentially with $t_{\pi\ell}-t_H$. This is a generic feature when calculating long-distance contributions in Euclidean space and such terms must be identified and subtracted. The number of these terms depends on $s_{\pi\ell}$ and on the chosen boundary conditions which in general will include \emph{twisting}. Note that no such exponentially growing terms are present for leptonic decays.

For $K_{\ell 3}$ decays, in some corners of phase space, there may also be multi-hadron intermediate states with energies smaller than the external one, and hence containing exponentials which grow with the time separation, but these are expected to be small. For example the $K\to\pi\pi\ell\nu\to\pi\ell\nu(\gamma)$ sequence only contributes at high order ($p^6$) in ChPT and is present due to the Wess-Zumino-Witten term in the action. More importantly however, we can restrict the values of $s_{\pi\ell}$ to a range below the multi-hadron threshold.
Note that for $D$ and $B$ decays the large number of such terms which need to be subtracted in most of phase space, makes it \emph{very difficult} to perform a non-perturbative lattice calculation.

\subsection{Finite-volume corrections} For leptonic decays of the pseudoscalar meson $P$, in QED$_\mathrm{L}$ finite-volume effects take the form:
\begin{equation}\Gamma_0^{\mathrm{pt}}(L) =  C_0(r_\ell) + \tilde C_0(r_\ell)\log\left(m_P L\right)+ \frac{C_1(r_\ell)}{m_P L}+ 
\dots \, ,\end{equation} where $r_\ell=m_\ell/m_P$\,\cite{Lubicz:2016xro}. An important point to note is that the exhibited $L$-dependent terms are \emph{universal}, i.e. independent of the structure of the meson and we have calculated these coefficients (using the QED$_{\mathrm{L}}$ regulator of the zero mode\,\cite{Hayakawa:2008an}). The leading structure-dependent FV effects in $\Gamma_0-\Gamma_0^{\mathrm{pt}}$ are of $O(1/L^2)$.

The following scaling law is useful in determining which terms need to be evaluated to obtain the universal coefficients. If the leading behaviour of the infinite-volume integrand and finite-volume summand is proportional to $1/(k^2)^{\hspace{-0.5pt}\frac n2}$ as $k\to 0$ then the corresponding difference between the infinite-volume integral and finite-volume sum of $O(1/L^{4-n})$\,\cite{Lubicz:2016xro}.
In the calculation of the mass spectrum $n=3$ and the leading finite-volume correction is of $O(1/L)$ and is universal, as is the subleading term of $O(1/L^2)$. In decay amplitudes $n=4$, corresponding to the presence of infrared divergences. 

\begin{figure}[t]
\begin{center}
\includegraphics[width=0.35\hsize]{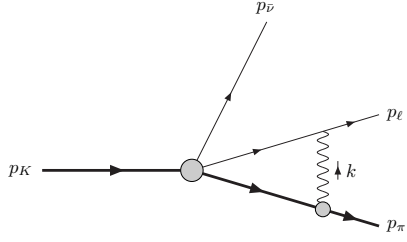}
\end{center}
\caption{Diagram contributing to the $K\to\pi\ell\bar{\nu}_\ell$ correlation function used in the discussion of finite-volume effects (see text). \label{fig:SL3}}
\end{figure}

For illustration consider the diagram in Fig.\,\ref{fig:SL3}. At small photon momentum $k$, the pion and lepton internal propagators scale as $1/k$ and the photon propagator as $1/k^2$, so that the loop integrand/summand scales as $1/k^4$ corresponding to an infrared divergence. There are also subleading terms which scale as $1/k^3$ which lead to $1/L$ finite-volume effects. These arise by expanding the propagators and vertices, including the vertex containing the weak Hamiltonian, to $O(k)$.
(Since the $1/L^2$ finite-volume corrections depend on the structure of the pion we do consider these further.)

\begin{figure}[t]
\begin{center}
\includegraphics[width=0.6\hsize]{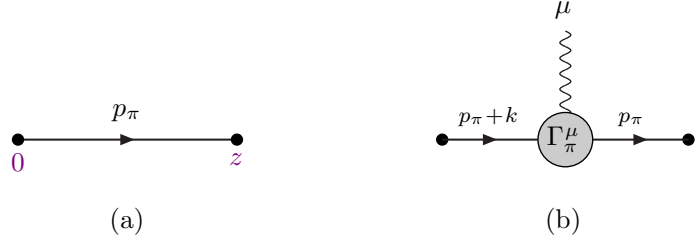}
\end{center}
\caption{(a) The pion propagator, (b) $\pi\gamma\pi$ vertex \label{fig:propandvertex}}
\end{figure}

Electromagnetic Ward identities are particularly useful in the study of the universality of the $O(1/L)$ finite-volume corrections. (Alternatively one can construct a gauge-invariant effective theory.) To illustrate this consider the pion propagator in Fig.\,\ref{fig:propandvertex}(a).
We define the Euclidean pion propagator $\Delta_\pi(p_\pi)$ by:
\begin{eqnarray}
C_{\pi\pi}(p_\pi)&=&\int d^{\,4}\!z~e^{-ip_\pi\cdot z}~\langle\,0\,|T\big\{\phi_\pi(z)
\phi^\dagger_\pi(0)\big\}\,|\,0\,\rangle\nonumber\\
&\equiv&\big|\langle\,0\,|\phi_\pi(0)\,|\,\pi(p_\pi)\,\rangle\big|^2~\Delta_\pi(p_\pi)\\ 
&\equiv&\big|\langle\,0\,|\phi_\pi(0)\,|\,\pi(p_\pi)\,\rangle\big|^2~\frac{Z_\pi(p_\pi^2)}{p_\pi^2+m_\pi^2}\,.\nonumber
\end{eqnarray}
$Z_\pi$ parametrises the structure dependence of the pion propagator. We now expand the propagator for small values of $k$ and off-shellness $\epsilon_\pi^2=p_\pi^2+m_\pi^2$ to obtain:
\begin{equation}
\Delta_\pi(p_\pi+k)=\frac{1-2z_{\pi_1}p_\pi\cdot k-\epsilon_\pi^2 z_{\pi_1}+O(k^2,\epsilon_\pi^4,\epsilon_\pi^2 k)}{\epsilon_\pi^2+2p_\pi\cdot k+k^2}\,,
\end{equation}
where the \underline{structure dependent} parameter $z_{\pi_1}$ is given by:
\begin{equation}
z_{\pi_1}=\left.\frac{dZ^{-1}_\pi(p_\pi^2)}{dp_\pi^2}\right|_{p_\pi^2=-m_\pi^2}\,,
\end{equation}
(the subscript {\footnotesize 1} on $z_{\pi_1}$ labels the coefficient of the Taylor series expansion of $Z_\pi^{-1}$\,\cite{Lubicz:2016xro}).

Similarly we define the amputated $\pi\gamma\pi$ vertex $\Gamma_\pi^\mu$, by amputating the propagators and matrix elements of the interpolating operators in the correlation function (see Fig.\,\ref{fig:propandvertex}(b))\\[-0.1in]
\begin{equation}
C_\pi^\mu(p_\pi,k)=i\int d^{\hspace{1pt}4}\hspace{-1pt}z\,d^{\hspace{1pt}4}\hspace{-1pt}x\,e^{-ip_\pi\cdot z}\,e^{-ik\cdot x}
\langle 0|\,T\big\{\phi_\pi(z)\,j^\mu(x)\,\phi_\pi^\dagger(0)\big\}\,|0\rangle
\,.\end{equation}
We now expand $\Gamma_\pi$ for small $k$ (and $\epsilon_\pi$).
The key result is obtained from the Ward identity:
\begin{equation}
k_\mu\Gamma^\mu_P(p_\pi,k)=Q_\pi\left\{\Delta_\pi^{-1}(p_\pi+k)-\Delta^{-1}_\pi(p_\pi)
\right\}\,,\end{equation}
which relates the first-order expansion coefficients and yields
\begin{equation}
Z_\pi(p_\pi+k)\,\Gamma_\pi^\mu(p_\pi,k)=Q_\pi\,(2p_\pi+k)^\mu+O(k^2,\epsilon_\pi^2)\,.\end{equation}
Here $Q_\pi$ is the electric charge of the pion. Thus, since we are neglecting the structure dependent $O(1/L^2)$ corrections, the pion propagator and $\pi\gamma\pi$ vertex combine to give the same result as in the point-like theory.

We have seen that, as a result of the Ward identity, we do not need the derivatives of the pion form-factors to obtain the $O(1/L)$ corrections. However, we also need to expand the weak-vertex which, in QCD without QED, is a linear combination of two form-factors $f^{\pm}(q^2)$. 
Off-shell, the $K\pi\ell\bar\nu$ weak vertex is a linear combination of two functions $F^{\pm}(p_\pi^2,p_K^2,2p_K\cdot p_\pi)$ (which on-shell reduce to the form-factors $f^{\pm}(q^2)$).
The Ward identity relates the $K\pi\ell\bar\nu$ and $K\pi\ell\bar\nu\gamma$ vertices and does lead to a partial, but not complete, cancellation of the $O(1/L)$ terms.
The remaining $O(1/L)$ corrections are found to depend on the derivatives of the form factors $df^{\pm}(q^2)/dq^2$, as well as on the form factors $f^{\pm}(q^2)$ themselves; this will be demonstrated in a publication in preparation.
Such derivative terms are a generic consequence of the Low theorem and are absent only in particularly  simple cases, such as leptonic decays as explained below.
These corrections are "universal" since the coefficients are physical, i.e. the form factors and their derivatives can be measured experimentally or computed in lattice simulations.
On the other hand, there are no corrections of the form $df^{\pm}\!/dm_\pi^2$ or $df^{\pm}\!/dm_K^2$, which would not be physical.

It is instructive to contrast the situation for semileptonic decays with the corresponding one for leptonic decays, e.g. \hspace{-11pt} for $K^+\to\ell^+\nu_\ell$ decays~\cite{Lubicz:2016xro}. In that case the leading isospin-breaking corrections are proportional to the decay constant $f_K$ computed in QCD simulations and again there are 
no $O(1/L)$ terms proportional to $df_K\!/dm_K^2$. In that case however, there is no scope for terms analogous to $df^{\pm}(q^2)/dq^2$.

For leptonic decays we had calculated the $O(1/L)$ finite-volume corrections  analytically using the Poisson summation formula\,\cite{Lubicz:2016xro}.
For semileptonic decays, we have calculated the integrands/summands necessary to evaluate the coefficients of the $O(1/L)$ corrections but have not yet evaluated the corrections themselves.
In the ignorance of the analytic coefficients, the subtraction of the $O(1/L)$ effects can be performed instead by fitting data obtained at different volumes with however, some loss of precision. For leptonic decays, where the $O(1/L)$ corrections are known and can be subtracted explicitly, we have checked that fitting these finite-volume effects numerically leads instead to an approximate doubling 
of the uncertainty in the theoretical prediction extrapolated to physical masses in the infinite volume limit. This may be disappointing, but recalling that isospin breaking corrections are of $O(1\%)$, it is not a major problem.

\section{The perturbative calculation}

We return now to the relation in Eq.\,(\ref{eq:SL1}) where we envisage that $d^2\Gamma_0^{\mathrm{pt}}/dq^2 ds_{\pi\ell}$ and \\
$d^2\Gamma_1/dq^2 ds_{\pi\ell}$ 
are to be calculated in perturbation theory.
This has not yet been fully done.

A related calculation has recently been performed by De Boer, Kitahari and Ni\v{s}and\v{z}i\'c\,\cite{deBoer:2018ipi} in the context of $B\to D^{(\ast)}$ semileptonic decays. 
This work was motivated by the $R(D)$ and $R(D^\ast)$ anomalies in semileptonic $B$-decays which seem to indicate a violation of lepton flavour universality between decays in which the final state charged lepton is a $\tau$ on the one-hand and a $\mu$ or electron on the other. 
The authors of Ref.\,\cite{deBoer:2018ipi} were investigating whether, within the Standard Model, this anomaly may be explained by 
radiative corrections not present in the \emph{photos} package; this appears not to be the case. The calculation however, is incomplete as we now explain.

The calculations in Ref.\,\cite{deBoer:2018ipi}  were not performed in the point-like approximation. Instead $d^2\Gamma_1/dq^2 ds_{\pi\ell}$ was obtained by using the eikonal approximation in which the denominators of the propagators of the charged lepton and meson are approximated by $\pm 2p\cdot k$, where $p$ is the momentum of the lepton or meson and $k$ is that of the photon. All powers of $k$ in the numerators (including at the weak vertex) are dropped. In the calculation of $d^2\Gamma_0^{\mathrm{pt}}/dq^2 ds_{\pi\ell}$, the dependence of the weak vertex on the photon's momentum $k$ is dropped, the form-factors are evaluated at the external value of $q^2$ (i.e. at $q=p_B-p_D$), but otherwise all factors of $k$ are kept\,\footnote{We thank Teppei Kitakara for helpful discussions on this point.}.

The formulae in Ref.\,\cite{deBoer:2018ipi} can be readily adapted to semileptonic kaon decays by changing the masses of the mesons and leptons. By inserting the results in Eq.\,(\ref{eq:SL1}) the infrared divergences cancel in both the terms on the right-hand separately. On the other hand, the fact that terms which behave as $1/k^3$ as $k\to 0$ are not fully evaluated implies that not all the $O(1/L)$ corrections are obtained. In particular, as explained above, we would need terms proportional to the derivative of the form factors.

\section{Summary and Conclusions}
We are developing the framework for the computation of radiative corrections  to semileptonic $K_{\ell 3}$ decays. This builds on the theoretical structure\,\cite{Carrasco:2015xwa,Lubicz:2016xro}, and its successful implementation\,\cite{Giusti:2017dwk,DiCarlo:2019thl}, developed for computations of radiative corrections to leptonic decays.
At this conference, we have also presented the results of a successful computation of  the $P\to\ell\bar\nu\gamma$ amplitude, making it possible to study leptonic decays of heavy mesons\,\cite{deDivitiis:2019uzm}.
Among the important points to note are:\\[0.03in]
(i) An appropriate observable to study for semileptonic decays is $d^2\Gamma/dq^2 ds_{\pi\ell}$.\\[0.02in]
(ii) The presence of exponentially growing terms in $t_{\pi\ell}-t_H$ which need to be subtracted. \\[0.02in]
(iii) The universality of the $O(1/L)$ corrections, which do however depend on the form-factors $f^{\pm}(q^2)$ and on their derivatives with respect to $q^2$. This is a
generic feature, absent only for particularly simple processes such as leptonic decays. (In the present study we have used the QED$_{\mathrm{L}}$
regulator for the photon's zero mode; similar techniques can be used to investigate the universality (or otherwise) of the $O(1/L)$ corrections using other regulators.)

\vspace{0.02in}Among the remaining things left to do is the analytic evaluation of the coefficients of the $O(1/L)$ corrections. Alternatively these corrections can be fitted numerically, in which case the result of Ref.\,\cite{deBoer:2018ipi} may be the most convenient one for the term which is added and subtracted in Eq.\,(\ref{eq:SL1}).
Finally the method needs to be implemented and tested numerically.\\[0.01in]

\vspace{-0.05in}\textbf{Acknowledgements:}
V.L., G.M. and S.S. thank MIUR (Italy) for partial support under the contract PRIN 2015. C.T.S. was supported by an Emeritus Fellowship from the Leverhulme Trust.
N.T. thanks the Univ. of Rome Tor Vergata for the support granted to the project PLNUGAMMA.


\end{document}